\documentclass[twocolumn,showpacs,superscriptaddress,10pt]{revtex4-1} 
\usepackage{amsmath,amssymb,graphicx}
\usepackage[T1]{fontenc}
\usepackage{bbold}
\usepackage{color}
\begin{document}

\title{Light propagation in biaxial crystals}

\author{Alex Turpin} \email{Corresponding author: alejandro.turpin@uab.cat}
\affiliation{Departament de F\'isica, Universitat Aut\`onoma de Barcelona, Bellaterra, E-08193, Spain}
\author{Yury V. Loiko}
\affiliation{Aston Institute of Photonic Technologies, School of Engineering and Applied Science Aston University, Birmingham, B4 7ET, UK}
\author{Todor K. Kalkandjiev}
\affiliation{Departament de F\'isica, Universitat Aut\`onoma de Barcelona, Bellaterra, E-08193, Spain}
\affiliation{Conerefringent Optics SL, Avda. Cubelles 28, Vilanova i la Geltr\'u, E-08800, Spain}
\author{Jordi Mompart}
\affiliation{Departament de F\'isica, Universitat Aut\`onoma de Barcelona, Bellaterra, E-08193, Spain}

\begin{abstract}
We present a formalism able to predict the transformation of light beams passing through biaxial crystals. We use this formalism to show both theoretically and experimentally the transition from double refraction to conical refraction, which is found when light propagates along one of the optic axes of a biaxial crystal. Additionally, we demonstrate that the theory is applicable both to non-cylindrically symmetric and non-homogeneously polarized beams by predicting the transformation of input beams passing through a cascade of biaxial crystals. 
\vspace{0.5cm}

\end{abstract}

\maketitle 

\section{Introduction}

Light propagation in optically anisotropic media is an interesting object of study due to the number of scientific and technological applications of uniaxial and biaxial crystals. Although the effects of anisotropic media on light beams is known since centuries ago, even nowadays this topic is an object of exhaustive study. Due to the fact that the mathematical description of uniaxial crystals is simpler than for the case of biaxial crystals, light propagation along the former is well understood. However, to our knowledge there is no basic set of equations offering the possibility of predicting the behavior of light after passing through biaxial crystals along any beam propagation direction. The most related works on that topic have been performed by Dreger \cite{dreger:1999:joa} and Garnier \cite{garnier:2001:jmp}. Analogously to uniaxial crystals, in biaxial crystals light propagating out of one of the optic axes suffers from double refraction. In contrast, if the beam propagates along one of the optic axes it suffers from conical refraction (CR), being this a phenomenon of renewed interest during the last years. In this case, the transformation of cylindrically symmetric and homogeneously polarized beams has been well described with diffractive optics by Belsky and Khapalyuk, and Berry \cite{belsky:1978a:os,berry:2004:jo}. 
Under conditions of CR, an input Gaussian beam with waist radius $w_0$, is transformed at the focal plane of the system into a pair of concentric bright rings split by a dark (Poggenforff) ring of geometrical radius $R_0$ provided that $\rho_0 \equiv R_0/w_0 \gg 1$ (see Fig.~\ref{fig2}(d)). No matter what the state of polarization (SOP) of the input beam is, at any point of the rings the SOP is linear with the azimuth rotating along the rings so that every two diametrically opposite points have orthogonal polarizations \cite{turpin_stokes:2014:oe}. Out of the focal plane, the beam evolves as an optical bottle with two on-axis maxima at positions $z = \sqrt{4/3}z_R \rho_0$ \cite{turpin_vault:2013:oe} (where $z_R$ is the Rayleigh length of the Gaussian input beam).

Up to know it has been shown both experimentally and theoretically CR for homogeneously polarized Gaussian input beams \cite{belsky:1978a:os,berry:2004:jo,kalkandjiev:2008:spie}, top-hat input beams \cite{dublin:2014:oe} and Laguerre-Gauss input beams \cite{peet:2011:ol,peet:2014:jo} through a single crystal and also the transformation of Gaussian input beams in cascaded crystals \cite{kalkandjiev:2008:spie,berry:2010:jo,amin:2011:jo,phelan:2012:oe,turpin_cascaded:2013:ol}. With a wave-vector based formalism, the CR transformation of elliptical input beams and an axicon input beam has been analyzed \cite{turpin_ebs:2013:oe} as well. Berry and Jeffrey also showed the transition from double refraction to CR for cylindrically symmetric beams with a modified treatment of CR that allows studying dichroic crystals \cite{berry:2006:jo}.  

Our aim here is to present a theoretical formalism capable to predict the transformation of light beams in biaxial crystals independently on the beam shape, its SOP and its propagation direction. We will provide the formulas for the electric field behind the crystal and show the evolution from double refraction to CR and the transformation of non-cylindrically symmetric and non-homogeneously polarized beams both in single and multiple biaxial crystals. As a proof of the usefulness of the theory, we will present the experimental cases of a Gaussian and an elliptical input beam.

\section{Theoretical formalism}
\label{formalism}
In what follows, we will consider normalized coordinates to $w_0$ and $z_R$, i.e. $x \rightarrow x/w_0$, $y \rightarrow y/w_0$ and $z \rightarrow z/z_R$. 
In the parabolic approximation, after passing through a medium or optical element, a light beam can be described by means of its displacement vector $\vec{D}$ as a superposition of plane waves $\vec{\kappa} = (\kappa_x,\kappa_y)$, which are generated from a unitary transformation provided by the optical element $\hat{U}(\vec{\kappa})$ applied over the Fourier transform vector of the input light beam $\vec{A}(\vec{\kappa})$. In other words,
\begin{equation}
\vec{D} = \frac{1}{(2 \pi)^2} \iint\limits_{-\infty}^{\infty} d\kappa_x d\kappa_y e^{i \vec{\kappa} \cdot \vec{r}} \hat{U}(\vec{\kappa}) \vec{A}(\vec{\kappa}),
\label{eq_D}
\end{equation}
where $\vec{A}(\vec{\kappa})$ is the Fourier transform 
\begin{eqnarray}
\vec{A}(\vec{\kappa}) &=& A_x(\vec{\kappa}) \vec{e_x} + A_y(\vec{\kappa}) \vec{e_y}, \\ \label{FT_vector}
A_{x}(\vec{\kappa}) &=& \frac{1}{(2 \pi)^2} \iint\limits_{-\infty}^{\infty}
E_{x}(\vec{r})  e^{-i \vec{\kappa} \cdot \vec{r}} dx dy,\\ \label{FT_x}
A_{y}(\vec{\kappa}) &=& \frac{1}{(2 \pi)^2} \iint\limits_{-\infty}^{\infty}
E_{y}(\vec{r})  e^{-i \vec{\kappa} \cdot \vec{r}} dx dy, \label{FT_y}
\end{eqnarray}
of the input beam with transverse electric field 
\begin{equation}
\vec{E}(\vec{r}) = E_x \vec{e_x} + E_y \vec{e_y},
\label{E_input}
\end{equation}
where $\vec{e_x}$ and $\vec{e_y}$ are the unitary vectors in Cartesian coordinates.

For a low birefringent biaxial crystal, it has been shown that the unitary transformation provided by the medium is $\hat{U}(\vec{\kappa}) = e^{-i \vec{\Gamma}(\vec{\kappa})}$ with \cite{berry:2004:jo,belsky:1978a:os,belsky:1978b:os}
\begin{equation}
\vec{\Gamma}(\vec{\kappa}) = \frac{1}{2} \kappa^2 z^2 \hat{\mathbf{I}} + \rho_0 \vec{\kappa} \cdot (\hat{\sigma_3},\hat{\sigma_1}),
\label{eq_gamma}
\end{equation}
where $\hat{\sigma_3}$ and $\hat{\sigma_1}$ are the Pauli matrices and $\hat{\mathbf{I}}$ is $2 \times 2$ the identity matrix. It is straightforward to demonstrate that for a given vector $\vec{v} = v \vec{n}$ with $|n| = 1$, the evaluation of $e^{i v (\vec{n} \cdot \vec{\sigma})}$ gives $\hat{\mathbf{I}} \cos(v) + i(\vec{n} \cdot \vec{\sigma} sin(v))$, where $\vec{\sigma} = (\hat{\sigma_1},\hat{\sigma_2},\hat{\sigma_3})$ is the vector of Pauli matrices. By recalling the latter identity and after evaluation of $\hat{U}(\vec{\kappa}) = e^{-i \vec{\Gamma}(\vec{\kappa})}$ by using Eq.~(\ref{eq_gamma}), the unitary transformation provided by the crystal can be obtained:
\begin{equation}
\hat{U}(\vec{\kappa}) = e^{-i \frac{1}{2} \kappa^2 z^2} \left[\cos(\rho_0 \kappa) \hat{\mathbf{I}} + i \frac{\sin(\rho_0 \kappa)}{\kappa} \vec{\kappa} \cdot (\hat{\sigma_3},\hat{\sigma_1}) \right],
\label{eq_U}
\end{equation}
where $\kappa \equiv \sqrt{\kappa_x^2 + \kappa_y^2}$.
By combining Eq.~(\ref{eq_D}) with Eq.~(\ref{eq_U}), there can be obtained two main integrals that describe the beam evolution behind the biaxial crystal: 
\begin{widetext}
\begin{eqnarray}
B_{0, \alpha}(\vec{r},\rho_0)&=& \frac{i}{(2 \pi)^2} \iint\limits_{-\infty}^{\infty} e^{-i (\vec{\kappa} \cdot \vec{r} - \frac{Z}{2n} \kappa^2))} \frac{\kappa_y}{\kappa} \sin \left( \rho_0 \kappa \right) A_{\alpha}(\vec{\kappa}) d \kappa_x d \kappa_y,\label{B0_CR} \\ 
B_{1, \alpha}(\vec{r},\rho_0) &=&
\frac{1}{(2 \pi)^2} \iint\limits_{-\infty}^{\infty} e^{-i (\vec{\kappa} \cdot \vec{r} - \frac{Z}{2n} \kappa^2)} \left( \cos \left( \rho_0 \kappa \right) + i \frac{\kappa_x}{\kappa} \sin \left( \rho_0 \kappa \right) \right) A_{\alpha}(\vec{\kappa}) d \kappa_x d \kappa_y,~\label{B1_CR}
\end{eqnarray}
\end{widetext}
where $\alpha=(x,y)$ and $n$ is the mean refractive index of the biaxial crystal. The expressions for the field $\vec{D}$ in terms of Eqs.~(\ref{B0_CR}) and (\ref{B1_CR}) are
\begin{eqnarray}
D_{x} &=& B_{0, y}(\vec{r},\rho_0) + B_{1, x}(\vec{r},\rho_0),
\label{E_CRx}\\
D_{y} &=& B_{0, x}(\vec{r},\rho_0) + B_{1, y}(\vec{r},-\rho_0). 
\label{E_CRy}
\end{eqnarray}

Note that Eqs.~(\ref{E_input})--(\ref{E_CRy}) do not require the input beam to be cylindrically symmetric nor to be homogeneously polarized. Additionally, since an unpolarized beam can be expressed as the incoherent addition of multiple beams, these equations can be used to predict the transformation of any input beam, no matter its SOP or its shape, as long as its Fourier transform can be obtained. For input beams with non-homogeneous polarizations, it must be taken into account that the beam and the polarization can be always decomposed in the $(x,y)$ basis, so that the theoretical formalism presented above can be always used. The above formalism reduces to the well known Belsky--Khapalyuk--Berry equations for CR when cylindrically symmetric and homogeneously polarized beams are considered, as demonstrated in Appendix~\ref{App:AppendixA}.

In what follows we will demonstrate the flexibility of the presented formalism to show the transformation of light beams in a variety of situations.

\section{Single crystal configuration}

In this section we discuss the transformation of input beams that propagate within a biaxial crystal non-parallel to one of the optic axes, i.e. under conditions of double refraction. We consider homogeneously left handed circularly polarized input beams, i.e with normalized electric field $\vec{e_0} = (1,i)/\sqrt{2}$, and we look at the transverse patterns at the focal plane ($Z=0$). Propagation out of the optic axis can be described by the product of the input field amplitude by a factor $\exp(-i \vec{\delta} \cdot \vec{r}_{\perp})$, where $\vec{\delta} = (\delta_x,\delta_y)$ and $\vec{r}_{\perp} = (x,y)$ \cite{berry:2006:jo}. 

We consider the well known case of an input beam with Gaussian transverse profile and an elliptical input beam. Their electric field amplitudes are described by 
\begin{eqnarray}
\vec{E}_{\rm{G}}(x,y) &=& e^{-i(\delta_x x+\delta_y y)} e^{-(x^2+y^2)}\vec{e_0}, \label{E_gauss} \\
\vec{E}_{\rm{EB}}(x,y) &=& e^{-i(\delta_x x+\delta_y y)} e^{-(\frac{x^2}{a^2}+\frac{y^2}{b^2})}\vec{e_0}, \label{E_EB} 
\end{eqnarray}
where $a$ and $b$ are constants and $\delta_x$ and $\delta_y$ give the angular separation of the input beam's propagation direction with respect to the optic axis of the crystal. 

\begin{figure}[htb]
\centering
\includegraphics[width=1 \columnwidth]{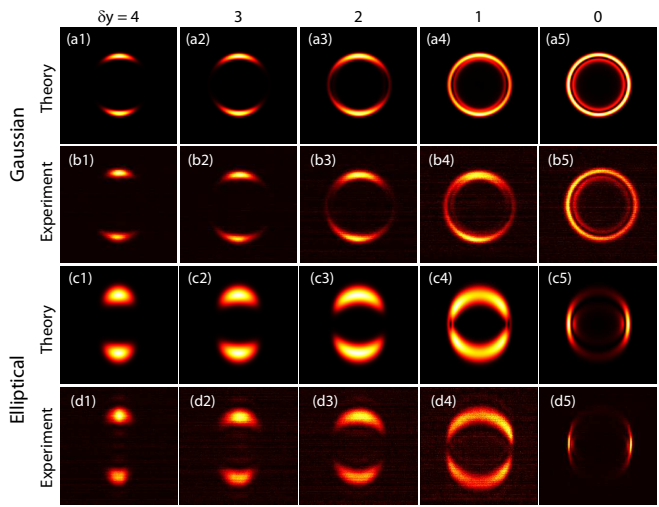}
\caption{Transverse intensity patterns showing the transition from double refraction to conical refraction for a Gaussian input beam (first and second rows) and an elliptical input beam (third and fourth rows) both experimentally (second and fourth rows) and numerically calculated by using Eqs.~(\ref{FT_vector})--(\ref{E_EB}) (first and third rows).}
\label{fig2}
\end{figure}

The experiments have been performed by employing a $\rm{KGd(WO_2)_4}$ (conicity $\alpha = 16.9\,\rm{mrad}$) biaxial crystal with a length of $l = 28\,\rm{mm}$, yielding a CR ring radius of $R_0 = l \alpha = 475\,\rm{\mu m}$. In Figs.\ref{fig2}(b1)--(b5) we used a Gaussian input beam focused by a spherical lens with $100\,\rm{mm}$ of focal length. The elliptical input beam was obtained by focusing the same Gaussian beam with a cylindrical lens with $100\,\rm{mm}$ of focal length. The biaxial crystal was mounted on an angular micrometric positioner that allowed to change the $\phi$ and $\theta$ angles in spherical coordinates and to observe the transition from double refraction to CR as the optic axis and the beam propagation direction approached each other. 

Fig.~\ref{fig2} shows transition from double refraction to CR for the Gaussian input beam (first and second rows) and the elliptical input beam (third and fourth rows) both experimentally (second and fourth rows) and numerically calculated by using Eqs.~(\ref{FT_vector})--(\ref{E_CRy}) (first and third  rows). For simplicity, we consider only angular displacement of the input beam in the vertical direction, i.e. $\delta_x = 0$.

The transformation of a Gaussian input beam propagating parallel to one of the optic axes of a biaxial crystal as described by Eq.~(\ref{E_gauss}) (parallel propagation implies $\delta_x=\delta_y=0$) is the most commonly studied case in CR. When $\delta_{x,y} \neq 0$ double refraction instead of CR is found. Double refraction in uniaxial crystals is associated with the splitting of the input beam into two beams with identical transverse pattern and orthogonal polarizations, corresponding to the ordinary and the extraordinary polarizations. However, in biaxial crystals, a Gaussian input beam propagating out of the optic axes splits into two azimuthal sectors placed at diametrically opposite positions of the otherwise expected CR ring, provided that the angular propagation deviation with respect to the optic axis is small. Only when $\delta_{x,y} \gg 1$ the output split beams preserve the input beam's pattern, as in uniaxial crystals. As the beam propagation direction approaches to the optic axis, the split beams occupy a larger azimuthal angle and eventually both interfere, see Figs.~\ref{fig2}(a3)--(a5). The fact that the two split beams interfere implies that both beams possess regions of non-orthogonal polarizations. For parallel propagation with respect to the optic axes, the interference between both split beams is maximum and the two bright rings with Poggendorff splitting are formed. 

In case of an elliptical input beam, there is a competition between the ellipticity of the shape of the beam and the refraction induced by the crystal. We consider an elliptical beam described by Eq.~(\ref{E_EB}) with $a=1, b=0.1$, i.e. with $w_x < w_y$. Since the misalignment with respect to the optic axis is only along the $y$ direction, the two split beams are expected also to appear in that direction. Due to the non-symmetrical nature of the elliptical beam and the double refraction provided by the biaxial crystal, which induce opposite effects, the two refracted beams for beam propagation out of the optic axis are wider than for the Gaussian input beam case, see Fig.~\ref{fig2}(c1). As the misalignment of the input beam is reduced, the refracted beams expand along the azimuthal direction and at some point both interfere, as in the previous case. For parallel propagation to one of the optic axis, the pattern is formed by two lobes each of which with Poggendorff splitting, see Fig.~\ref{fig2}(c5). The two lobes are slightly connected between each other but we have checked that the connection points tend to disappear as the ratio of the axes of the ellipse increases. We have additionally checked that there is a continuous evolution of the double-concentric ring structure from Fig.~\ref{fig2}(a5) into the double-lobe pattern from Fig.~\ref{fig2}(c5) as the ratio of the axes of the ellipse increases. For a deeper study of CR from elliptical beams, see Ref.~\cite{turpin_ebs:2013:oe}. 

\section{Cascade of crystals}
\begin{figure}[htb]
\centering
\includegraphics[width=0.8\columnwidth]{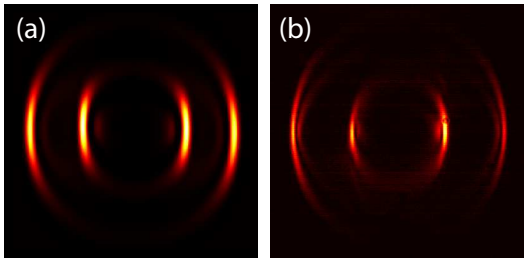}
\caption{Transverse intensity pattern at the focal plane for a cascade of two biaxial crystals with aligned optic axes obtained from an elliptical input beam. (a) Numerical simulations obtained by using Eqs.~(\ref{B0_CR})--(\ref{E_EB}). (b) Experimental measurements obtained by using two $\rm{KGd(WO_2)_4}$ biaxial crystals with lengths of $l_1 = 28\,\rm{mm}$ and $l_2 = 10\,\rm{mm}$.}
\label{fig3}
\end{figure}

Light propagation through a cascade of crystals can be predicted by using Eqs.~(\ref{E_input})--(\ref{E_CRy}) recursively. In this case, the output electric field after the first crystal $\vec{E}_{\rm{CR:1}}$ becomes the input electric field impinging the second biaxial crystal and this process can be repeated until passing through the whole cascade. Since the beams obtained after the first crystal can be non-cylindrically symmetric and non-homogeneously polarized, the results reported in what follows prove also the usefulness of our formulation for such input beams.
Fig.~\ref{fig3} presents the numerical simulations and experimental results obtained for a cascade of two biaxial crystal with common optic axis being parallel to the optical axis of the system for an elliptical input beam. The passage of an elliptical input beam through the two crystal cascade shows a similar behavior: each of the two lobes with Poggendorff splitting obtained after the first crystal split into two lobes, also with Poggendorff splitting, after passing through the second biaxial crystal. The experimental observations from Fig.~\ref{fig3}(b) were performed by using two $\rm{KGd(WO_2)_4}$ biaxial crystals with lengths of $l_1 = 28\,\rm{mm}$ and $l_2 = 10\,\rm{mm}$. The Gaussian and elliptical input beams were identical to the ones used in the previous section. As it can be appreciated, both theory and experiment agree well.

\section{Conclusions}
In summary, we have presented a theoretical formalism that describes the beam evolution of light propagating through biaxial crystals. The method can be used with complete generality, i.e. it does not require symmetries in the beam shape nor in its state of polarization, as well as can be used to predict the behavior of light with any propagation direction. We have demonstrated the flexibility of the reported formalism for the transformation of a wide variety of beams, including non-cylindrically symmetric and non-homogeneously polarized input beams, propagating along one of the optic axes of a biaxial crystal. Additionally, we have shown the transition from double to conical refraction for non-cylindrically symmetric beams, such as an elliptical beam. Finally, we have reported the case of a cascade of multiple biaxial crystals. Our numerical predictions accurately reproduce previous experiments on light transformation in biaxial crystals \cite{kalkandjiev:2008:spie,peet:2014:jo,dublin:2014:oe,turpin_ebs:2013:oe}.

We expect the reported results to be applied in all areas of optics requiring the use of biaxial crystals and, in particular, in optical trapping \cite{turpin_vault:2013:oe,phelan:2010:oe,edik:2014:ol,turpin_PDR:2015:oe}, free space optical communications \cite{turpin:2012:ol}, polarimetry \cite{peinado:2013:ol,peinado:2015:oe}, lasing and non-linear optical phenomena \cite{loiko:2014:SPIE,marc:2014:ol}, laser processing \cite{ciaran:2011:oc}, the generation and manipulation of vector beams \cite{turpin_stokes:2014:oe} and beam shaping \cite{peet:2010:oe,loiko:2013:ol,turpin:2014:ol}.

\section*{Acknowledgments}
The authors gratefully acknowledge financial support through the Spanish Ministry of Science and Innovation (MICINN) (contract FIS2011-23719) and the Catalan Government (contract SGR2014-1639). A.T. acknowledges financial support from the MICINN through the grant AP2010-2310.

\appendix
\section{Cylindrically symmetric solution} 
\label{App:AppendixA}
In this Appendix we will demonstrate that the general formalism presented in Section~\ref{formalism} can be used to derive the fundamental Belsky--Khapalyuk--Berry equations of conical refraction for input beams possessing both cylindrical symmetry and homogeneous polarization.
By using Eq.~(\ref{eq_U}), Eq.~(\ref{eq_D}) can be written in Cartesian coordinates as follows: 
\begin{widetext}
\begin{equation}
\vec{D} = \frac{1}{(2 \pi)^2} \iint\limits_{-\infty}^{\infty} e^{i \vec{\kappa} \cdot \vec{r}} e^{-i \frac{1}{2 n} \kappa^2 Z^2} \left[\cos(\rho_0 \kappa) \hat{\mathbf{I}} + i \frac{\sin(\rho_0 \kappa)}{\kappa} \vec{\kappa} \cdot (\hat{\sigma_3},\hat{\sigma_1}) \right] \vec{A}(\vec{\kappa}) d\kappa_x d\kappa_y.
\label{eq_D_1}
\end{equation}
\end{widetext}
The electric field described by Eq.~(\ref{eq_D_1}) can be rewritten in terms of two fundamental equations as shown by Eqs.~(\ref{B0_CR})--(\ref{E_CRy}). 

If we change the coordinate system from Cartesian to cylindrical, we must carry out the following replacements: 
\begin{eqnarray}
\vec{\kappa} &\leftrightarrow& \kappa (\cos(\phi_{\kappa}), \sin(\phi_{\kappa})), \\ 
\vec{r} &\leftrightarrow& \rho (\cos(\varphi), \sin(\varphi)),\\
e^{i \vec{\kappa} \cdot \vec{r}} &\leftrightarrow& e^{i \kappa \rho \cos(\phi_{\kappa} - \varphi)}, \\
\iint\limits_{-\infty}^{\infty} d \kappa_x d \kappa_y &\leftrightarrow& \int_{0}^{\infty} \kappa d\kappa \int_{0}^{2 \pi} d\phi_{\kappa},
\end{eqnarray}
where we consider the radial coordinate to be normalized to the beam waist radius ($w_0$) of the input beam through $\rho \equiv r/w_0$.
In this case, Eq.~(\ref{eq_D_1}) reads as follows:
\begin{widetext}
\begin{equation}
\vec{D} = \frac{1}{(2 \pi)^2} \int_{0}^{\infty} \int_{0}^{2 \pi} e^{i k \rho \cos(\phi_{\kappa} - \varphi)} e^{-i \frac{1}{2 n} \kappa^2 Z^2}
\left[
\cos (\rho_0 \kappa)
\left( \begin{array}{cc}
1 & 0 \\
0 & 1 \\
 \end{array} \right)
+i \sin (\rho_0 \kappa)
\left( \begin{array}{cc}
\cos(\phi_{\kappa}) & \sin(\phi_{\kappa}) \\
\sin(\phi_{\kappa}) & -\cos(\phi_{\kappa}) \\
 \end{array} \right)
\right]
\vec{A}(\vec{\kappa}) \kappa d\kappa d\phi_{\kappa},
\label{eq_D_2}
\end{equation}
\end{widetext}
Now, we simplify the system by considering a uniformly polarized and cylindrically symmetric input beam $\vec{E}~=~E(r)~\vec{e_0}$ with cylindrically symmetric Fourier transform too, i.e. $A_x(\vec{\kappa}) = A_y(\vec{\kappa}) = a(\kappa)$, where 
\begin{equation}
a(\kappa) = \int\limits_{0}^{\infty} \kappa E(r) J_0(r \kappa) dr~,
\label{FT_cyl}
\end{equation}
being $J_n (x)$ the n$^{\rm{th}}$-order Bessel function of the first type. 
With this assumption, we can integrate over $\phi_{\kappa}$ by using the following expressions
\begin{eqnarray}
\int\limits_{0}^{2 \pi} e^{i \kappa \rho \cos(\phi_{\kappa} - \varphi)} d\kappa &=& J_0(k \rho)~,\label{aux1}\\
\int\limits_{0}^{2 \pi} e^{i \kappa \rho \cos(\phi_{\kappa} - \varphi)} \cos(\phi_{\kappa}) d\kappa &=& \cos(\varphi) J_1(k \rho)~\label{aux2} \\
\int\limits_{0}^{2 \pi} e^{i \kappa \rho \cos(\phi_{\kappa} - \varphi)} \sin (\phi_{\kappa}) d\kappa &=& 0 ~\label{aux3}
\end{eqnarray}
and obtain a 1D integral for the $\vec{D}$ field:  
\begin{equation}
\vec{D} = 
\left( \begin{array}{cc}
B_0 + B_1 \cos(\varphi) & B_1 \sin(\varphi) \\
B_1 \sin(\varphi) & B_0 - B_1 \cos(\varphi) \\
 \end{array} \right) \vec{e_0},
\label{eq_D_2}
\end{equation}
where $B_0 = B_0(\rho,Z,\rho_0)$ and $B_1 = B_1(\rho,Z,\rho_0)$ are the Belsky--Khapalyuk--Berry integrals that describe the evolution of the CR beam both in the radial and axial directions:
\begin{eqnarray}
B_0(r,Z,\rho_0) =\frac{1}{(2 \pi)} \int\limits_{0}^{\infty} a(\kappa) e^{-i \frac{\kappa^2 Z^2}{2 n}}  \cos(\kappa \rho_0) J_0(\kappa r) d\kappa~,\label{B0}\\
B_0(r,Z,\rho_0) = \frac{1}{(2 \pi)} \int\limits_{0}^{\infty} a(\kappa) e^{-i \frac{\kappa^2 Z^2}{2 n}}  \sin(\kappa \rho_0) J_1(\kappa r) d\kappa~.\label{B1}
\end{eqnarray}


\begin{thebibliography}{1}

\bibitem{dreger:1999:joa}
M. A. Dreger, 
``Optical beam propagation in biaxial crystals,''
 J. Opt. A: Pure Appl. Opt. \textbf{1}, 601--616 (1999).

\bibitem{garnier:2001:jmp}
J. Garnier, 
``High-frequency asymptotics for Maxwell's equations in anisotropic media Part I: Linear geometric and diffractive optics,''
J. Math. Phys. \textbf{42}, 1612--1635 (2001).

\bibitem{belsky:1978a:os} 
A. M. Belskii and A. P. Khapalyuk, 
``Internal conical refraction of bounded light beams in biaxial crystals,'' 
Opt. Spectrosc. \textbf{44}, 436--439 (1978). 

\bibitem{berry:2004:jo}
M. V. Berry,
``Conical diffraction asymptotics: fine structure of Poggendorff rings and axial spike,'' 
J. Opt. A. \textbf{6}, 289–-300 (2004).

\bibitem{turpin_stokes:2014:oe}
A. Turpin, Yu. V. Loiko, A. Peinado, A. Lizana, J. Campos, T. K. Kalkandjiev, and J. Mompart,
``Polarization tailored novel vector beams based on conical refraction,''
Opt. Express \textbf{23}, 5704--5715 (2015).

\bibitem{turpin_vault:2013:oe}
A. Turpin, V. Shvedov, C. Hnatovsky, Yu. V. Loiko, J. Mompart, and W. Krolikowski, 
``Optical vault: A reconfigurable bottle beam based on conical refraction of light,'' 
Opt. Express \textbf{21}, 26335--26340 (2013).

\bibitem{kalkandjiev:2008:spie} T. K. Kalkandjiev and M. A. Bursukova, ``Conical refraction: an experimental introduction,'' in \textit{Photon Management III}, J. T. Sheridan, F. Wyrowski, eds., Proc. SPIE \textbf{6994}, 69940B--69940B-10 (2008).

\bibitem{dublin:2014:oe}
R. T. Darcy, D. McCloskey, K. E. Ballantine, J. G. Lunney, P. R. Eastham, and J. F. Donegan, 
``Conical diffraction intensity profiles generated using a top-hat input beam,'' 
Opt. Express \textbf{22}, 11290--11300 (2014).

\bibitem{peet:2011:ol}
V. Peet, 
``Conical refraction and formation of multiring focal image with Laguerre--Gauss light beams,'' 
Opt. Lett. \textbf{36}, 2913--2915 (2011).

\bibitem{peet:2014:jo}
V. Peet, 
``Experimental study of internal conical refraction in a biaxial crystal with Laguerre–Gauss light beams,'' 
J. Opt. \textbf{16}, 075702 (2014).

\bibitem{berry:2010:jo}
M. V. Berry, 
``Conical diffraction from an N-crystal cascade,'' 
J. Opt. \textbf{12}, 075704 (2010).

\bibitem{amin:2011:jo}
A. Abdolvand, ``Conical diffraction from a multi-crystal cascade: experimental observations,'' 
Appl. Phys. B \textbf{103}, 281--283 (2011).

\bibitem{phelan:2012:oe}
C. F. Phelan, K. E. Ballantine, P. R. Eastham, J. F. Donegan, and J. G. Lunney, 
``Conical diffraction of a Gaussian beam with a two crystal cascade,'' 
Opt. Express \textbf{20}, 13201--13207 (2012). 

\bibitem{turpin_cascaded:2013:ol}
A. Turpin, Yu. V. Loiko, T. K. Kalkandjiev, and J. Mompart, 
``Multiple rings formation in cascaded conical refraction,'' 
Opt. Lett. \textbf{38}, 1455--1457 (2013).

\bibitem{turpin_ebs:2013:oe}
A. Turpin, Yu. V. Loiko, T. K. Kalkandjiev, H. Tomizawa, and J. Mompart, 
``Wave-vector and polarization dependence of conical refraction,'' 
Opt. Express \textbf{21}, 4503--4511 (2013).

\bibitem{berry:2006:jo}
M. V. Berry and M. R. Jeffrey, 
``Conical diffraction complexified: dichroism and the transition to double refraction,'' 
J. Opt. A, Pure Appl. Opt. \textbf{8}, 1043--1051 (2006).

\bibitem{belsky:1978b:os} A. M. Belskii and A. P. Khapalyuk, 
``Propagation of confined light beams along the beam axes (axes of single ray velocity) of biaxial crystals,'' 
Opt. Spectrosc. \textbf{44}, 312--315 (1978).

\bibitem{phelan:2010:oe}
D. P. O'Dwyer, C. F. Phelan, K. E. Ballantine, Y. P. Rakovich, J. G. Lunney, and J. F. Donegan, 
``Conical diffraction of linearly polarised light controls the angular position of a microscopic object,'' 
Opt. Express \textbf{18}, 27319--27326 (2010).

\bibitem{edik:2014:ol}
C. McDonald, C. McDougall, E. Rafailov, and D. McGloin, 
``Characterizing conical refraction optical tweezers,'' 
Opt. Lett. \textbf{39}, 6691--6694 (2014).

\bibitem{turpin_PDR:2015:oe}
A. Turpin, J. Polo, Yu. V. Loiko, J. K\"uber, F. Schmaltz, T. K. Kalkandjiev, V. Ahufinger, G. Birkl, and J. Mompart,
``Blue-detuned optical ring trap for Bose-Einstein condensates based on conical refraction,''
Opt. Express \textbf{23}, 1638--1650 (2015).

\bibitem{turpin:2012:ol} A. Turpin, Yu. V. Loiko, T. K. Kalkandjiev and J. Mompart, 
``Free-space optical polarization demultiplexing and multiplexing by means of conical refraction,'' Opt. Lett. \textbf{37}, 4197--4199 (2012).

\bibitem{peinado:2013:ol}
A. Peinado, A. Turpin, A. Lizana, E. Fernández, J. Mompart, and J. Campos, 
``Conical refraction as a tool for polarization metrology,''
Opt. Lett. \textbf{38}, 4100--4103 (2013).

\bibitem{peinado:2015:oe}
A. Peinado, A. Lizana, A. Turpin, C. Iemmi, T. K. Kalkandjiev, J. Mompart, and J. Campos, 
``Optimization, tolerance analysis and implementation of a Stokes polarimeter based on the conical refraction phenomenon,''
Opt. Express \textbf{23}, 5636-5652 (2015).

\bibitem{loiko:2014:SPIE}
Y. V. Loiko, G. S. Sokolovskii, D. Carnegie, A. Turpin, J. Mompart, and E. U. Rafailov, 
``Laser beams with conical refraction patterns''
Proc. SPIE \textbf{8960}, 89601Q (2014).

\bibitem{marc:2014:ol}
R. Cattoor, I. Manek-H\"onninger, D. Rytz, L. Canioni, and M. Eichhorn, 
``Laser action along and near the optic axis of a holmium-doped KY(WO$_4$)$_2$ crystal,'' 
Opt. Lett. \textbf{39}, 6407--6410 (2014).

\bibitem{ciaran:2011:oc}
C. F. Phelan, R. J. Winfield, D. P. O'dwyer, Y. P. Rakovich, J. F. Donegan, and J. G. Lunney,
``Two-photon polymerisation of novel shapes using a conically diffracted femtosecond laser beam,''
Opt. Commun \textbf{284}, 3571--3574 (2011).

\bibitem{peet:2010:oe}
V. Peet, 
``Improving directivity of laser beams by employing the effect of conical refraction in biaxial crystals,'' 
Opt. Express \textbf{18}, 19566--19573 (2010).

\bibitem{loiko:2013:ol}
Yu. V. Loiko, A. Turpin, T. K. Kalkandjiev, E. U. Rafailov, and J. Mompart, 
``Generating a three-dimensional dark focus from a single conically refracted light beam,'' 
Opt. Lett. \textbf{38}, 4648--4651 (2013).

\bibitem{turpin:2014:ol}
A. Turpin, Yu. V. Loiko, T. K. Kalkandkiev, H. Tomizawa, and J. Mompart, 
``Super-Gaussian conical refraction beam,'' 
Opt. Lett. \textbf{39}, 4349--4352 (2014).

\end{thebibliography}
\end{document}